\documentclass[aps,prb,onecolumn]{revtex4}
\usepackage{epsfig}
\usepackage{amsmath}
\usepackage{amsfonts}
\usepackage{amssymb}
\usepackage{graphicx}
\usepackage{colordvi}

\begin{document}


\title{DARK ENERGY MODELS TOWARD OBSERVATIONAL TESTS AND DATA}

\author{SALVATORE CAPOZZIELLO}

\address{Dipartimento di Scienze Fisiche, Universit\`a di Napoli "Federico II" and INFN, Sez. di Napoli, \\
Complesso Universitario di Monte S. Angelo Ed.G, Via Cinthia,
I-80126 - Napoli, Italy \\
capozziello@na.infn.it }


\begin{abstract}
A huge amount of  good quality astrophysical data converges
towards the picture of a spatially flat universe undergoing the
today observed phase of accelerated expansion. This new
observational trend is commonly addressed as {\it Precision
Cosmology}. Despite of the quality of astrophysical surveys, the
nature of  dark energy dominating the matter-energy content of the
universe is still unknown and a lot of different scenarios are
viable candidates to explain cosmic acceleration. Methods  to test
these cosmological models are based on distance measurements and
lookback time toward  astronomical objects used as standard
candles. I discuss  the characterizing parameters and constraints
of three different classes of dark energy  models pointing out the
related degeneracy problem which is the signal  that  more data at
low $(z\sim 0\div 1)$, medium $(1<z<10)$ and high $(10 <z< 1000)$
redshift are needed to definitively select realistic  models.
\end{abstract}

\maketitle


\section{Introduction}

The increasing bulk of data that have been accumulated  in the
last few years have paved the way to the emergence of a new
standard cosmological model usually referred to as the {\it
concordance model}. The Hubble diagram of Type Ia Supernovae
(hereafter SNeIa), measured by both the Supernova Cosmology
Project \cite{SCP} and the High\,-\,z Team \cite{HZT} up to
redshift $z \sim 1$, was the first evidence  that the universe is
undergoing a phase of accelerated expansion. On the other hand,
balloon born experiments, such as BOOMERanG \cite{Boomerang} and
MAXIMA \cite{Maxima}, determined the location of the first and
second peak in the anisotropy spectrum of  cosmic microwave
background radiation (CMBR) strongly pointing out that the
geometry of the universe is spatially flat. If combined with
constraints coming from galaxy clusters on the matter density
parameter $\Omega_M$, these data indicate that the universe is
dominated by a non-clustered fluid with negative pressure,
generically dubbed {\it dark energy}, which is able to drive the
accelerated expansion. This picture has been further strengthened
by the more precise measurements of the CMBR spectrum, due to the
WMAP experiment \cite{WMAP}, and by the extension of the SNeIa
Hubble diagram to redshifts higher than 1 \cite{Riess04}. After
these observational evidences, an  overwhelming flood of papers,
presenting a great variety of models trying to explain this
phenomenon, has appeared; in any case,  the simplest explanation
is claiming for the well known cosmological constant $\Lambda$
\cite{LCDMrev}. Although the best fit to most of the available
astrophysical data \cite{WMAP}, the $\Lambda$CDM model failed in
explaining why the inferred value of $\Lambda$ is so tiny (120
orders of magnitude lower) compared to the typical vacuum energy
values predicted by particle physics and why its energy density is
today comparable to the matter density  (the so called {\it
coincidence problem}). As a tentative solution, many authors have
replaced the cosmological constant with a scalar field rolling
down its potential and giving rise to the model now referred to as
{\it quintessence} \cite{QuintRev}. Even if successful in fitting
the data, the quintessence approach to dark energy is still
plagued by the coincidence problem since the dark energy and
matter densities evolve differently and reach comparable values
for a very limited portion of the universe evolution  coinciding
at present era. In this case, the coincidence problem is replaced
with a fine-tuning problem.  Moreover, it is not clear where this
scalar field originates from thus leaving a great uncertainty on
the choice of the scalar field potential. The subtle and elusive
nature of  dark energy has led many authors to  look for
completely different scenarios able to give a quintessential
behavior without the need of exotic components. To this aim, it is
worth stressing that the acceleration of the universe only claims
for a negative pressure dominant component, but does not tell
anything about the nature and the number of cosmic fluids filling
the universe. This consideration suggests that it could be
possible to explain the accelerated expansion by introducing a
single cosmic fluid with an equation of state causing it to act
like dark matter at high densities and dark energy at low
densities. An attractive feature of these models, usually referred
to as {\it Unified Dark Energy} (UDE) or {\it Unified Dark Matter}
(UDM) models, is that such an approach naturally solves, al least
phenomenologically, the coincidence problem. Some interesting
examples are the generalized Chaplygin gas \cite{Chaplygin}, the
tachyon field \cite{tachyon} and the condensate cosmology
\cite{Bassett}. A different class of UDE models has been proposed
\cite{Hobbit} where a single fluid is considered whose energy
density scales with the redshift in such a way that the radiation
dominated era, the matter domination era and the accelerating
phase can be naturally achieved. It is worth noting that such
class of models are extremely versatile since they can be
interpreted both in the framework of UDE models and as a two-fluid
scenario with dark matter and scalar field dark energy. The main
ingredient of the approach is that a generalized equation of state
can be always obtained and observational data can be fitted.
Actually, there is still a different way to face the problem of
cosmic acceleration. As  stressed in Lue et al. \cite{LSS03}, it
is possible that the observed acceleration is not the
manifestation of another ingredient in the cosmic pie, but rather
the first signal of a breakdown of our understanding of the laws
of gravitation. From this point of view, it is thus tempting to
modify the Friedmann equations to see whether it is possible to
fit the astrophysical data with  models comprising only the
standard matter. Interesting examples of this kind are the
Cardassian expansion \cite{Cardassian} and the DGP gravity
\cite{DGP}. Moving in this same framework, it is possible to  find
alternative schemes where a quintessential behavior is obtained by
 taking into account effective models coming from fundamental physics
giving rise to generalized or higher order gravity actions
\cite{curvature}. For instance, a cosmological constant term may
be recovered as a consequence of a non\,-\,vanishing torsion field
thus leading to a model which is consistent with both SNeIa Hubble
diagram and Sunyaev\,-\,Zel'dovich data coming from clusters of
galaxies \cite{torsion}. SNeIa data could also be efficiently
fitted including higher-order curvature invariants in the gravity
Lagrangian \cite{curvfit}. It is worth noting that these
alternative schemes provide naturally a cosmological component
with negative pressure whose origin is  related to the geometry of
the universe thus overcoming the problems linked to the physical
significance of the scalar field. It is evident, from this short
overview, the high number of cosmological models which are viable
candidates to explain the observed accelerated expansion. This
abundance of models is from one hand the signal of the fact that
we have a  limited number of cosmological tests to discriminate
among rival theories, and from the other hand that a urgent
degeneracy problem has to be faced.  To this aim, it is useful to
remember that both the SNeIa Hubble diagram and the angular
size\,-\,redshift relation of compact radio sources \cite{AngTest}
are distance based methods to probe cosmological models so then
systematic errors and biases could be iterated. From this point of
view, it is interesting to look for tests based on time-dependent
observables. For example, one can take into account the {\it
lookback time} to distant objects since this quantity can
discriminate among different cosmological models. The lookback
time is observationally estimated as the difference between the
present day age of the universe and the age of a given object at
redshift $z$. Such an estimate is possible if the object is a
galaxy observed in more than one photometric band since its color
is determined by its age as a consequence of stellar evolution. It
is thus possible to get an estimate of the galaxy age by measuring
its magnitude in different bands and then using stellar
evolutionary codes to choose the model that reproduces the
observed colors at best. A quite similar approach was pursued by
Lima \& Alcaniz \cite{LA00} who used the age (rather than the
lookback time) of old high redshift galaxies to constrain the dark
energy equation of state (see also \cite{Jimenez}). The same
method was then applied also to braneworld models \cite{AJD02} and
the Chaplygin gas \cite{AJD03}. It is worth noting, however, that
the estimate of the  age of a single galaxy may be affected by
systematic errors which are difficult to control. Actually, this
problem can be overcome by considering a sample of galaxies
belonging to the same cluster. In this way, by averaging the
estimates of all galaxies, one obtains an estimate of the cluster
age and reduces the systematic errors. Such a method was first
proposed by Dalal et al. \cite{DAJM01} and then used by Ferreras
et al. \cite{FMT03} to test a class of models where a scalar field
is coupled with the matter term giving rise to a particular
quintessence scheme. These lectures are devoted to review
different classes of dark energy models discussing some methods to
constrain them toward observational data. Far from being
exhaustive and complete, my aim is to point out the degeneracy
problem and the fact that we  need further and self-consistent
observational surveys at {\it all} redshifts to remove it. The
layout of these notes are the  following. In Sect.\,2, I briefly
present  three classes of cosmological models pointing out the
main quantities which we  need for distance measurements and
lookback time tests. Methods to fit and constrain models are
discussed in Sect.\,3. In Sect.\,4, I take into account a sample
of galaxy clusters as an example of data useful to fit dark energy
models at low redshifts. As exercise, in Sect.\,5, the above
classes of models are fitted against the sample of galaxy clusters
considering also SNeIa and WMAP data and results. The aim of this
section is pedagogical: the goal is to show the emergence of the
degeneracy problem. Summary and conclusions are given in Sect.\,6.

\section{Dark energy models}

As discussed in the Introduction, many rival theories  have been
advocated to fit the observed accelerated expansion and to solve
the puzzle of the nature of the dark energy. As a simple
classification scheme , we may divide the different cosmological
models in three wide classes. According to the models of the first
class, the dark energy is a new ingredient of the cosmic Hubble
flow, the simplest case being the $\Lambda$CDM scenario and its
quintessential generalization which we will refer to as QCDM
models. This is in sharp contrast with the assumption of UDE
models (the second class) where there is a single fluid described
by an equation of state comprehensive of all regimes of cosmic
evolution  \cite{Hobbit} which I will consider here referring to
it as the {\it parametric density models} or generalized {\it
EoS}\footnote{EoS for Equation of State.} models. Finally,
according to the third class models, accelerated expansion is the
first evidence of a breakdown of the Einstein General Relativity
(and thus of the Friedmann equations) which has to be considered
as a particular case of a more general theory of gravity. As an
example of this kind of models, we will consider the
$f(R)$\,-\,gravity \cite{curvature,curvfit}. Far from being
exhaustive, considering these three classes of models allow to
explore very different scenarios proposed to explain the observed
cosmic acceleration. In the following we will sketch these  three
dark energy approaches  and derive  some of the main quantities
which we need for matching the observations.

\subsection{The $\Lambda$CDM model and its  generalization toward quintessence (QCDM models)}

\begin{figure}
\centering \resizebox{6.5cm}{!}{\includegraphics{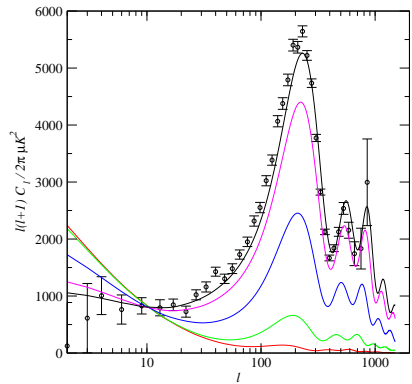}}
\caption{The CMBR anisotropy spectrum for different values of $w$.
Data points are the WMAP measurements and the best fit is obtained
for $w\simeq -1$.} \label{fig: zqcl}
\end{figure}

Cosmological constant $\Lambda$ has  become  a textbook candidate
to drive the accelerated expansion of the spatially flat universe.
Despite its {\it conceptual} problems, the $\Lambda$CDM model
turns out to be the best fit to a combined analysis of completely
different astrophysical data ranging from SNeIa to CMBR anisotropy
spectrum and galaxy clustering \cite{WMAP,SDSS03,VSA}. As a simple
generalization, one may consider the QCDM scenario in which the
barotropic factor $w \equiv p/\rho$ takes at a certain epoch a
negative value with $w = -1$ corresponding to the standard
cosmological constant. Testing whether  such a barotropic factor
deviate or not from $-1$ is one of the main issue of modern
observational cosmology. How such a negative pressure fluid drives
the cosmic acceleration may be easily understood by looking at the
Friedmann equations\,:

\begin{equation}
H^2 \equiv \left ( \frac{\dot{a}}{a} \right )^2 = \frac{8 \pi G}{3} (\rho_{M} + \rho_Q) \ ,
\label{eq: fried1}
\end{equation}

\begin{equation}
2 \frac{\ddot{a}}{a} + H^2 = - 8 \pi G p_Q = - 8 \pi G w \rho_Q \ ,
\label{eq: fried2}
\end{equation}
where the dot denotes differentiation with respect to  cosmic time
$t$, $H$ is the Hubble parameter and the universe is assumed
spatially flat as suggested by the position of the first peak in
the CMBR anisotropy spectrum  (see also Fig.\ref{fig: zqcl}).
)\cite{Boomerang,Maxima,WMAP}.

From the continuity equation, $\dot{\rho} + 3 H (\rho + p) = 0$,
we get for the $i$\,-\,th fluid with $p_i = w_i \rho_i$\,:

\begin{equation}
\Omega_i = \Omega_{i,0} a^{-3 (1 + w_i)} = \Omega_{i,0} (1 + z)^{3 (1 + w_i)} \ ,
\label{eq: omegavsz}
\end{equation}
where $z \equiv 1/a - 1$ is the redshift, $\Omega_i =
\rho_i/\rho_{crit}$  is the density parameter for the $i$\,-\,th
fluid in terms of the critical density $\rho_{crit} = 3H_0^2/8\pi
G$ and, hereafter, I label all the quantities evaluated today with
a subscript $0$. Inserting this result into Eq.(\ref{eq: fried1}),
one gets\,:

\begin{equation}
H(z) = H_0 \sqrt{\Omega_{M,0} (1 + z)^3 + \Omega_{Q,0} (1 + z)^{3 (1 + w)}} \ .
\label{eq: hvsz}
\end{equation}
Using Eqs.(\ref{eq: fried1}), (\ref{eq: fried2}) and the
definition of the deceleration parameter $q \equiv - a
\ddot{a}/\dot{a}^2$, one finds\,:

\begin{equation}
q_0 = \frac{1}{2} + \frac{3}{2} w (1 - \Omega_{M,0}) \ .
\label{eq: qlambda}
\end{equation}
The SNeIa Hubble diagram, the large scale galaxy clustering  and
the CMBR anisotropy spectrum can all be fitted by the $\Lambda$CDM
model with $(\Omega_{M,0}, \Omega_Q) \simeq (0.3, 0.7)$ thus
giving $q_0 \simeq -0.55$, i.e. the universe turns out to be in an
accelerated expansion phase. The simplicity of the model and its
capability of fitting the most of the data are the reasons why the
$\Lambda$CDM scenario is the leading candidate to explain the dark
energy cosmology. Nonetheless,  its generalization,  QCDM models,
i.e.  mechanisms allowing the evolution of $\Lambda$ from the past
are invoked  to remove the $\Lambda$-problem and the $coincidence$
problem.

\subsection{Generalizing the EoS: the parametric density models}

In the framework of UDE models, it has been  proposed
\cite{Hobbit} a phenomenological class of models by introducing a
single fluid\footnote{It is worth stressing that this model may be
interpreted not only as comprising a single fluid with an exotic
equation of state, but also as made of dark matter and scalar
field dark energy or in the framework of modified Friedmann
equations. For pedagogical reasons, I prefer the UDE
interpretation even if the results do not depend on this
assumption.} with energy density\,:

\begin{equation}
\rho(a) = A_{norm} \left ( 1 + \frac{s}{a} \right )^{\beta - \alpha} \
\left [ 1 + \left ( \frac{b}{a} \right )^{\alpha} \right ]
\label{eq: rhor}
\end{equation}
with $0 < \alpha < \beta$, $s$ and $b$ (with $s < b$) two  scaling
factors and $A_{norm}$ a normalization constant. For several
applications, it is useful to rewrite the energy density as a
function of the redshift $z$. Replacing $a = (1 + z)^{-1}$ in
Eq.(\ref{eq: rhor}), we get\,:

\begin{equation}
\rho(z) = A_{norm} \ \left ( 1 + \frac{1 + z}{1 + z_s} \right )^{\beta - \alpha} \
\left [ 1 + \left ( \frac{1 + z}{1 + z_b} \right )^{\alpha} \right ]
\label{eq: rhoz}
\end{equation}
having defined $z_s = 1/s - 1$ and $z_b = 1/b - 1$.   It is easy
to show that $\rho \propto a^{-\beta}$ for $a << s$, $\rho \propto
a^{-\alpha}$ for $s << a << b$ and $\rho \propto const.$ for $a >>
b$. By setting $(\alpha, \beta) = (3, 4)$ the energy density
smoothly interpolates from a radiation dominated phase to a matter
dominated period finally approaching a de Sitter phase. The
normalization constant $A_{norm}$ may be estimated by inserting
Eq.(\ref{eq: rhor}) into Eq.(\ref{eq: fried1}) and evaluating the
result today. This gives\,:

\begin{equation}
A_{norm} = \frac{\rho_{crit,0}}{(1 + s)^{\beta - \alpha} \ (1 + b^{\alpha})} \ .
\label{eq: norm}
\end{equation}
The continuity equation may be recast in a form that allows   to
compute the pressure and  the barotropic factor $w = p/\rho$
obtaining \cite{Hobbit}\,:

\begin{equation}
w = \frac{[ (\alpha - 3) a + (\beta - 3) s] b^{\alpha} - [3 (a + s) + (\alpha - \beta) s ] a^{\alpha}}{3 (a + s) (a^{\alpha} + b^{\alpha})}
\label{eq: wr}
\end{equation}
which shows that the barotropic factor strongly depends on  the
scale factor (and hence on the redshift). Combining the Friedmann
equations, the deceleration parameter $q = (1 + 3 w)/2$ is

\begin{equation}
q = \frac{[(\alpha - 2) a + (\beta - 2) s] b^{\alpha} - [2 (a + s) + (\alpha - \beta) s] a^{\alpha}}
{2 (a + s) (a^{\alpha} + b^{\alpha})} \ ;
\label{eq: qr}
\end{equation}
inserting $a = 1$ gives the present day value, i.e.:

\begin{equation}
q_0 = \frac{(y - 1) \alpha + z_s [\alpha \ y - 2 (1 + y)] + (\beta - 4) (1 + y)}{2 (2 + z_s) (1 + y)}
\label{eq: qz}
\end{equation}
with $y = (1 + z_b)^{-\alpha}$.  Some  straightforward
considerations allows to derive the following constraints on $q_0$
\cite{Hobbit}\,:

\begin{equation}
\frac{1}{2} \left [ \frac{\beta - \alpha}{2 + z_s} - 2 \right ] \le q_0
\le \frac{1}{2} \left [ \frac{\alpha z_s + 2 \beta}{2 (2 + z_s)} - 2 \right ] \ .
\label{eq: qlimits}
\end{equation}
It is convenient to solve Eq.(\ref{eq: qz}) with respect to  $z_b$
in order to express this one as a function of $q_0$ and $z_s$. It
is\,:

\begin{equation}
z_b = \left [ \frac{\alpha (1 + z_s) + \beta - (2 + z_s) (2 q_0 + 2)}{\alpha - \beta + (2 + z_s) (2 q_0 + 2)} \right ]^{1/\alpha} - 1 \ .
\label{eq: solvezb}
\end{equation}
Such a parametric density model is fully characterized  by five
parameters which are chosen to be the two asymptotic slopes
$(\alpha, \beta)$, the present day values of the deceleration
parameter and of the Hubble constant $(q_0, H_0)$ and the scaling
redshift $z_s$. As in \cite{Hobbit}, we will set $(\alpha, \beta)
= (3, 4)$ and $z_s = 3454$ so that $(q_0, H_0)$ will be the two
parameters to be constrained by the data. Any generalized EoS
approach can be reduced to this scheme which is useful to fit the
data. However, the phenomenological parameters, can often have a
fundamental physics counterpart.

\subsection{Extended Gravity: the case of curvature quintessence}

There is no {\it a priori} reason to restrict  the gravity action
to a linear function of Ricci scalar $R$ as in the General
Relativity. Actually, higher order curvature invariants always
emerge in quantum field theory formulated on curved spacetimes so
 it is worth considering the effect of such generalizations also on
the late evolution of the universe.

Replacing the Ricci scalar $R$ with a (up to now) generic
function $f(R)$ of Ricci scalar in the gravity lagrangian, the
resulting field equations may still be recast in the
Friedmann\,-\,like form provided that the total energy density and
the total pressure are written as\,:

\begin{displaymath}
\rho_{tot} = \rho_M + \rho_{curv} \ \ \ \ , \ \ \ \ p_{tot} = p_M + p_{curv}
\end{displaymath}
where the energy density and the pressure due to the
higher\,-\,order  curvature invariants are \cite{curvature}\,:

\begin{equation}
\rho_{curv} = \frac{1}{f^{\prime}(R)} \left \{ \frac{1}{2} \left [
f(R) - R f^{\prime}(R) \right ] - 3 H \dot{R} f^{\prime \prime}(R)
\right \} \ , \label{eq: rhocurv}
\end{equation}

\begin{eqnarray}
p_{curv} & = \displaystyle{\frac{1}{f^{\prime}(R)}}  & \left \{ (2
H \dot{R} + \ddot{R}) f^{\prime \prime}(R) +
\dot{R}^2 f^{\prime \prime \prime}(R) \right . \nonumber \\
~ & ~ & \ \ \left . + \frac{1}{2} \left [ f(R) - R f^{\prime}(R) \right ] \right \}  \ .
\label{eq: pcurv}
\end{eqnarray}
This approach is particularly useful since it allows to  interpret
the non-Einstein part of  gravitational interaction as a {\it new
fluid} with energy density and pressure given by (\ref{eq:
rhocurv}) and (\ref{eq: pcurv}) respectively. The barotropic
factor for such a fluid turns out to be\,:

\begin{equation}
w_{curv} = -1 + \frac{f^{\prime \prime}(R) \ddot{R} + \left [
f^{\prime \prime \prime}(R)  \dot{R} - H f^{\prime \prime}(R)
\right ]} {\left [ f(R) - R f^{\prime}(R) \right ]/2 - 3 H
f^{\prime \prime}(R)} \ . \label{eq: wcurv}
\end{equation}
A leading role is played by the form chosen for $f(R)$. As a
simple choice \cite{curvature}, we set $f(R) = f_0 R^n$ with $f_0$
constant.  It is possible to find power law solutions of the
Friedmann equations as $a(t) = (t/t_0)^{\alpha}$ with:

\begin{equation}
\alpha = \frac{2 n^2 - 3 n +1}{2 - n} \ .
\label{eq: alphancurv}
\end{equation}
This gives a class of cosmological models with constant
deceleration parameter\,:

\begin{equation}
q(t) = q_0 = \frac{1 - \alpha}{\alpha} = - \frac{2 n^2 - 2 n - 1}{2 n^2 - 3 n + 1} \ .
\label{eq: qzcurv}
\end{equation}
In order to get an accelerated expansion ($\alpha > 0$ and $q_0 <
0$),  the parameter $n$ has to satisfy the following constraint\,:

\begin{equation}
n \in \left [ -\infty, \frac{1 - \sqrt{3}}{2} \right ] \cup \left [ \frac{1 + \sqrt{3}}{2}, \infty \right ] \ .
\label{eq: nranges}
\end{equation}
I will refer to models with $n$ in the first (second) range as
{\it CurvDown} ({\it CurvUp}) models respectively and use the
method described in the following section to constrain the two
parameters $(n, H_0)$ which completely assign the model. A comment
is necessary at this point. Actually a simple power-law
$f(R)$-gravity model is not sufficient to realistically reproduce
a matter-dominated transient phase (needed for large scale
structure formation) evolving into the today observed accelerated
phase \cite{CNOT}. However, also if  more realistic models are
needed \cite{mimick}, it is useful to show that accelerated
behaviour can be recovered simply by extending General Relativity
without introducing further exotic ingredients (see also the
contributions by Francaviglia,  Nojiri, Odintsov and Schmidt in
these Lecture Notes).

\section{Methods to constrain models}
In this section, I will discuss how cosmological models can be
constrained, in principle, using suitable distance and/or time
indicators. As a general remark,  solutions coming from
cosmological models have to be matched with observations by using
the redshift $z$ as the natural time variable for the Hubble
parameter, i.e.

\begin{equation} H(z)=-\frac{\dot{z}}{z+1}\,. \end{equation}  Interesting
ranges for $z$  are: $100< z < 1000$ for early universe (CMBR
data), $10 < z < 100$ (LSS), $0 < z < 10$ (SNeIa, radio-galaxies,
etc.). The method consists in building up a reasonable patchwork
of data coming from different epochs and then
 matching them with the same cosmological solution ranging, in principle, from
inflation to present accelerated era. In order to constrain the
parameters characterizing the cosmological solution, a reasonable
approach is to maximize the following likelihood function\,:

\begin{equation}
{\cal{L}} \propto \exp{\left [ - \frac{\chi^2({\bf p})}{2} \right
]} \label{eq: deflike1}
\end{equation}
where {\bf p} are the parameters characterizing  the cosmological
solution. The $\chi^2$ merit function can be defined as\,:

\begin{equation}
\chi^2({\bf p})  =  \sum_{i = 1}^{N}{\left [ \frac{y^{th}(z_i,
{\bf p}) - y_i^{obs}}{\sigma_i} \right ]^2}  +
\displaystyle{\left [ \frac{{\cal{R}}({\bf p}) - 1.716}{0.062}
\right ]^2} + \displaystyle{\left [ \frac{{\cal{A}}({\bf p}) -
0.469}{0.017} \right ]^2}  \ . \label{eq: defchi1}\
\end{equation}

Terms entering Eq.(\ref{eq: defchi1}) can be characterized as
follows. For example, the dimensionless coordinate distances $y$
to objects at redshifts $z$ are considered in the first term. They
are defined as\,:

\begin{equation}
y(z) = \int_{0}^{z}{\frac{dz'}{E(z')}} \label{eq: defy}
\end{equation}
where $E(z)=H(z)/H_0$ is the normalized Hubble parameter. This is
the main quantity which allows to compare the theoretical results
with data. The function $y$ is related to the luminosity distance
$D_L = (1 + z) r(z)$.  A sample of data on $y(z)$ for the 157
SNeIa is discussed in the Riess et al. \cite{Riess04} Gold dataset
and 20 radio-galaxies are in \cite{DD04}. These authors fit with
good accuracy the linear Hubble law at low redshift ($z < 0.1$)
obtaining the Hubble dimensionless parameter $h = 0.664 {\pm}
0.008\,.$  Such a number can be consistently taken into account at
low redshift.  This value  is in agreement with $H_0 = 72 {\pm} 8
\ {\rm km \ s^{-1} \ Mpc^{-1}}$ given by the HST Key project
\cite{Freedman} based on the local distance ladder and estimates
coming from  time delays in multiply imaged quasars \cite{H0lens}
and  Sunyaev\,-\,Zel'dovich effect in X\,-\,ray emitting clusters
\cite{H0SZ}. The second term in Eq.(\ref{eq: defchi1}) allows to
extend the $z$-range to probe $y(z)$ up to the last scattering
surface $(z\geq 1000)$.  The {\it shift parameter}
\cite{WM04,WT04} $ {\cal R} \equiv \sqrt{\Omega_M} y(z_{ls}) $ can
be determined from the CMBR anisotropy spectrum, where $z_{ls}$ is
the redshift of the last scattering surface which can be
approximated as  $ z_{ls} = 1048 \left ( 1 + 0.00124
\omega_b^{-0.738} \right ) \left ( 1 + g_1 \omega_M^{g_2} \right )
$ with $\omega_i = \Omega_i h^2$ (with $i = b, M$ for baryons and
total matter respectively) and $(g_1, g_2)$ given in \cite{HS96}.
The parameter $\omega_b$ is constrained by the baryogenesis
calculations contrasted to the observed abundances of primordial
elements. Using this method, the value $ \omega_b = 0.0214 {\pm}
0.0020  $  is found \cite{Kirk}. In any case, it is worth noting
that the exact value of $z_{ls}$ has a negligible impact on the
results and setting $z_{ls} = 1100$ does not change constraints
and priors on the other  parameters of the given model. The third
term in the function $\chi^2$ takes into account  the {\it
acoustic peak} of the large scale correlation function at $100 \
h^{-1} \ {\rm Mpc}$ separation, detected by using  46748 luminous
red galaxies (LRG) selected from the SDSS Main Sample
\cite{Eis05,SDSSMain}. The quantity

\begin{equation}
{\cal{A}} = \frac{\sqrt{\Omega_M}}{z_{LRG}} \left [
\frac{z_{LRG}}{E(z_{LRG})} y^2(z_{LRG}) \right ]^{1/3} \label{eq:
defapar}
\end{equation}
is related to the position of acoustic peak where $z_{LRG} = 0.35$
is the effective redshift of the above sample. The parameter
${\cal{A}}$ depends  on the dimensionless coordinate distance (and
thus on the integrated expansion rate),  on $\Omega_M$ and $E(z)$.
This dependence removes some of the degeneracies intrinsic in
distance fitting methods. Due to this reason, it is particularly
interesting to include ${\cal{A}}$ as a further constraint on the
model parameters using its measured value  $ {\cal{A}} = 0.469
{\pm} 0.017  $  \cite{Eis05}. Note that, although similar to the
usual $\chi^2$ introduced in statistics, the reduced $\chi^2$
(i.e., the ratio between the $\chi^2$ and the number of degrees of
freedom) is not forced to be 1 for the best fit model because of
the presence of the priors on ${\cal{R}}$ and ${\cal{A}}$ and
since the uncertainties $\sigma_i$ are not Gaussian distributed,
but take care of both statistical errors and systematic
uncertainties. With the definition (\ref{eq: deflike1}) of the
likelihood function, the best fit model parameters are those that
maximize ${\cal{L}}({\bf p})$.

 Using the method sketched above, the classes of models studied here can be
constrained and selected by  observations. However, most of the
tests recently used to constrain cosmological parameters (such as
the SNeIa Hubble diagram and the angular size\,-\,redshift) are
essentially distance\,-\,based methods. The proposal of Dalal et
al. \cite{DAJM01} to use the lookback time to high redshift
objects is thus particularly interesting since it relies on a
completely different observable. The lookback time is defined as
the difference between the present day age of the universe and its
age at redshift $z$ and may be computed as\,:

\begin{equation}
t_L(z, {\bf p}) = t_H \int_{0}^{z}{\frac{dz'}{(1 + z') E(z', {\bf p})}}
\label{eq: deftl}
\end{equation}
where $t_H = 1/H_0 = 9.78 h^{-1} \ {\rm Gyr}$ is the Hubble  time
(with $h$ the Hubble constant in units of $100 \ {\rm km \ s^{-1}
\ Mpc^{-1}}$),  and, as above, $E(z, {\bf p}) = H(z)/H_0$ is the
dimensionless Hubble parameter and  $\{{\bf p}\}$ the set of
parameters characterizing the cosmological model. It is worth
noting that, by definition, the lookback time is not sensible to
the present day age of the universe $t_0$ so that it is (at least
in principle) possible that a model fits well the data on the
lookback time, but nonetheless it predicts a completely wrong
value for $t_0$. This latter parameter can be evaluated from
Eq.(\ref{eq: deftl}) by simply changing the upper integration
limit from $z$ to infinity. This shows that it is indeed a
different quantity since it depends on the full evolution of the
universe and not only on how the universe evolves from the
redshift $z$ to now. That is why this quantity can be explicitly
introduced as a further constraint. As an example, let us now
discuss how to use the lookback time and the age of the universe
to test a given cosmological model. To this end, let us consider
an object $i$ at redshift $z$ and denote by $t_i(z)$ its age
defined as the difference between the age of the universe when the
object was born, i.e. at the formation redshift $z_F$, and the one
at $z$. It is\,:

\begin{eqnarray}
t_i(z) & = & \displaystyle{\int_{z}^{\infty}{\frac{dz'}{(1 + z')
E(z', {\bf p})}} - \int_{z_F}^{\infty}{\frac{dz'}{(1 + z') E(z', {\bf p})}}} \nonumber \\
~ & = & \displaystyle{\int_{z}^{z_F}{\frac{dz'}{(1 + z') E(z',
{\bf p})}}}
 =  t_L(z_F) - t_L(z) \ . \label{eq: titl}
\end{eqnarray}
where, in the last row, we have used the definition (\ref {eq:
deftl})  of the lookback time. Suppose now we have $N$ objects and
we have  been able to estimate the age $t_i$ of the object at
redshift $z_i$ for $i = 1, 2, \ldots, N$. Using the previous
relation, we can estimate the lookback time $t_{L}^{obs}(z_i)$
as\,:

\begin{eqnarray}
t_{L}^{obs}(z_i) & = & t_L(z_F) - t_i(z) \nonumber \\
~ & = & [t_0^{obs} - t_i(z)] - [t_0^{obs} - t_L(z_F)] \nonumber \\
~ & = & t_{0}^{obs} - t_i(z) - df \ ,
\label{eq: deftlobs}
\end{eqnarray}
where $t_{0}^{obs}$ is the today estimated age of  the universe
and a {\it delay factor} can be defined as\,:

\begin{equation}
df = t_0^{obs} - t_L(z_F) \ .
\end{equation}
The delay factor is introduced to take into account  our ignorance
of the formation redshift $z_F$ of the object. Actually, what can
be measured is the age $t_i(z)$ of the object at redshift $z$. To
estimate $z_F$, one should use Eq.(\ref{eq: titl}) assuming a
background cosmological model. Since our aim is to determine what
is the background cosmological model, it is clear that we cannot
infer $z_F$ from the measured age so that this quantity is
completely undetermined. It is worth stressing that, in principle,
$df$ should be different for each object in the sample unless
there is a theoretical reason to assume the same redshift at the
formation of all the objects. If this is indeed the case (as I
will assume later), then it is computationally convenient to
consider $df$ rather than $z_F$ as the unknown parameter to be
determined from the data. Again a likelihood function can be
defined as\,:

\begin{equation}
{\cal{L}}_{lt}({\bf p}, df) \propto \exp{[-\chi^{2}_{lt}({\bf p}, df)/2]}
\label{eq: deflikelt}
\end{equation}
with\,:

\begin{equation}
\chi^{2}_{lt} = \displaystyle{\frac{1}{N - N_p + 1}} \left \{
\left [ \frac{t_{0}^{theor}({\bf p}) -
t_{0}^{obs}}{\sigma_{t_{0}^{obs}}} \right ]^2
 + \sum_{i = 1}^{N}{\left [ \frac{t_{L}^{theor}(z_i, {\bf p}) -
t_{L}^{obs}(z_i)}{\sqrt{\sigma_{i}^2 + \sigma_{t}^{2}}} \right
]^2} \right \} \label{eq: defchi}
\end{equation}
where $N_p$ is the number of parameters of the model,  $\sigma_t$
is the uncertainty on $t_{0}^{obs}$, $\sigma_i$ the one on
$t_{L}^{obs}(z_i)$ and the superscript {\it theor} denotes the
predicted values of a given quantity. Note that the delay factor
enters the definition of $\chi^2_{lt}$ since it determines
$t_{L}^{obs}(z_i)$ from $t_i(z)$ in virtue of Eq.(\ref{eq:
deftlobs}), but the theoretical lookback time does not depend on
$df$. In principle, such a method should work efficiently to
discriminate among the various dark energy models. Actually, this
is not exactly the case due to the paucity of the available data
which leads to large uncertainties on the estimated parameters. In
order to partially alleviate this problem, it is convenient to add
further constraints on the models by using  Gaussian
priors\footnote{The need for  priors to reduce the parameter
uncertainties is often advocated for  cosmological tests. For
instance, in Ref.\,\cite{LA00} a strong prior on $\Omega_M$ is
introduced to  constrain the dark energy equation of state. It is
likely, that extending the dataset to higher redshifts and
reducing the uncertainties on the age estimate will allow to avoid
resorting to priors.} on the Hubble constant, i.e. redefining the
likelihood function as\,:

\begin{equation}
{\cal{L}}({\bf p}) \propto {\cal{L}}_{lt}({\bf p}) \exp{\left [
-\frac{1}{2} \left ( \frac{h - h^{obs}}{\sigma_h}  \right )^2
\right ]} \propto \exp{[- \chi^2({\bf p})/2]} \label{eq: deflike}
\end{equation}
where we have absorbed $df$ in the set of parameters ${\bf p}$ and have defined\,:
\begin{equation}
\chi^2 = \chi^{2}_{lt} + \left ( \frac{h - h^{obs}}{\sigma_h} \right )^2
\label{eq: newchi}
\end{equation}
with $h^{obs}$ the estimated value of $h$ and  $\sigma_h$ its
uncertainty. The HST Key project results \cite{Freedman} can be
used setting $(h, \sigma_h) = (0.72, 0.08)$. Note that this
estimate is independent of the cosmological model since it has
been obtained from local distance ladder methods. The best fit
model parameters ${\bf p}$ may  be obtained by maximizing
${\cal{L}}({\bf p})$ which is equivalent to minimize the $\chi^2$
defined in Eq.(\ref{eq: newchi}). It is worth stressing that such
a function should not be considered as a {\it statistical
$\chi^2$} in the sense that it is not forced to be of order 1 for
the best fit model to consider a fit as successful. Actually, such
an interpretation is not possible since the errors on the measured
quantities (both $t_i$ and $t_0$) are not Gaussian distributed
and, moreover, there are uncontrolled systematic uncertainties
that may also dominate the error budget. Nonetheless, a
qualitative comparison among different models may be obtained by
comparing the values of this pseudo $\chi^2$ even if this should
not be considered as a definitive evidence against a given model.
Having more than one parameter, one obtains the best fit  value of
each single parameter $p_i$ as the value which maximizes the
marginalized likelihood for that parameter defined as\,:

\begin{equation}
{\cal{L}}_{p_i} \propto \int{dp_1 \ldots
\int{dp_{i - 1} \int{dp_{i + 1} \ldots \int{dp_n \ {\cal{L}}({\bf p})}}}} \ .
\label{eq: deflikemar}
\end{equation}
After having normalized  the marginalized likelihood to 1  at
maximum, one computes the $1 \sigma$ and $2 \sigma$ confidence
limits (CL) on that parameter by solving ${\cal{L}}_{p_i} =
\exp{(-0.5)}$ and ${\cal{L}}_{p_i} = \exp{(-2)}$ respectively. In
summary, taking into account the above procedures for distance and
time measurements, one can reasonably constrain a given
cosmological model. In any case, the main and obvious issue is to
have at disposal sufficient and good quality data sets.

\section{Samples of data to constrain models: the case of LSS}

In order to apply the method outlined above, we need a  set of
distant objects whose age can be somehow estimated. For example,
clusters of galaxies seem to be ideal candidates in this sense
since they can be detected up to high redshift and their redshift,
at formation epoch \footnote{It is worth stressing that, in
literature, the cluster formation redshift is defined as the
redshift at which the last episode of star formation happened. In
this sense, we should modify our definition of $df$ by adding a
constant term which takes care of how long is the star formation
process and what is the time elapsed from the beginning of the
universe to the birth of the first cluster of galaxies. For this
reason, it is still possible to consider the delay factor to be
the same for all clusters, but it is not possible to infer $z_F$
from the fitted value of $df$ because we do not know the detail of
 star formation history. This approach is
particular useful since it allows  to overcome the problem to
consider lower limits of the universe age at $z$ rather than  the
actual values.} is almost the same for all the clusters.
Furthermore, it is relatively easy to estimate their age from
photometric data only. To this end, the color of their component
galaxies,  in particular the reddest ones,  is needed. Actually,
the stellar populations of the reddest galaxies become redder and
redder as they evolve. It is just a matter, then, to assume a
stellar population synthesis model, and to look at how old the
latest episode of star formation should be happened in the galaxy
past to produce colors as red as the observed ones. This is what
is referred to as {\it color age}. The main limitation of the
method relies in the stellar population synthesis model, and on a
few (unknown) ingredients (among which the metallicity and the
star formation rate). The choice of the evolutionary model is a
key step  in the estimate of the color age and the main source of
uncertainty \cite{Worthey}. An alternative and more robust route
to cluster age is to consider the color scatter (see \cite{Bower}
for an early application of this approach). The argument,
qualitatively, goes in this way\,: if galaxies have an extreme
similarity in their color and nothing is conspiring to make the
color scatter surreptitiously small, then the latest episode of
star formation should happen in the galaxy far past, otherwise the
observed color scatter would be larger. Quantitatively, the
scatter in color should thus be equal to  the derivative of color
with time multiplied the scatter of star formation times. The
first quantity may be predicted using population synthesis models
and turns out to be almost the same for all the evolutionary
models thus significantly reducing the systematic uncertainty. We
will refer to the age estimated by this method as {\it scatter
age}. The dataset we need to apply the method described in the
previous section may now be obtained using the following
procedure. First, for a given redshift $z_i$, we collect the
colors of the redddest galaxies in a cluster at that redshift and
then use one of the two methods outlined above to determine the
color or the scatter age of the cluster. If more than one cluster
is available at that redshift, we average the results from
different clusters in order to reduce systematic errors. Having
thus obtained $t_i(z_i)$, we then use Eq.(\ref{eq: deftlobs}) to
estimate the value of the lookback time at that redshift.
Actually, what we measure is $t_{L}^{obs}(z_i) + df$ that is the
quantity that enters the definition (\ref{eq: defchi}) of
$\chi^2_{lt}$ and then the likelihood function. To estimate the
color age, following \cite{Andreon4},  it is possible to choose,
among the various available stellar population synthesis models,
the Kodama and Arimoto one \cite{Kodama1}, which, unlike other
models, allows a chemical evolution neglected elsewhere. This
gives us three points on the diagram $z$ vs. $t_{L}^{obs}$
obtained by applying the method to a set of six clusters at three
different redshifts as detailed in Table\,1. Using a large sample
of low redshift SDSS clusters, it is possible to evaluate the
scatter age for clusters age at $z = 0.10$ and $z = 0.25$
\cite{Andreon1}.  Blakeslee et al. \cite{Blakeslee} applied the
same method to a single, high redshift $(z = 1.27)$ cluster.
Collecting the data using both  the color age and the scatter age,
we end up with a sample of $\sim 160$ clusters at six redshifts
(listed in Table\,1) which probe the redshift range $(0.10,
1.27)$. This nicely overlaps the one probed by SNeIa Hubble
diagram so that a comparison among our results and those from
SNeIa is possible. I assume a $\sigma = 1 \ {\rm Gyr}$ as
uncertainty on the cluster age, no matter what is the method used
to get that estimate. Note that this is a very conservative
choice. Actually, if the error on the age were so large, the
color\,-\,magitude relation for reddest cluster galaxies should
have a large scatter that is not observed. I have, however, chosen
such a large error to take qualitatively into account the
systematic uncertainties related to the choice of the evolutionary
model.

\begin{table}
\begin{center}
\begin{tabular}{|c|c|c|c|c|c|c|c|}
\hline
 \multicolumn{4}{|c|}{Color age} & \multicolumn{4}{|c|}{Scatter age} \\
\hline $z$  & N & Age (Gyr) & Ref & $z$  & N & Age (Gyr) & Ref \\
\hline\hline
0.60 & 1 & 4.53 & \cite{Andreon4} & 0.10 & 55 & 10.65 & \cite{Andreon1} \\
0.70 & 3 & 3.93 & \cite{Andreon4} & 0.25 & 103 & 8.89 & \cite{Andreon1} \\
0.80 & 2 & 3.41 & \cite{Andreon4} & 1.27 &   1 & 1.60 & \cite{Blakeslee} \\
\hline
\end{tabular}
\end{center}
\caption{Main properties of the cluster sample used  for  the
analysis. The data in the left part of the Table refers to
clusters whose age has been estimated from the color of the
reddest galaxies (color age), while that of clusters in the right
part has been obtained by the color scatter (scatter age). For
each data point, I give the redshift $z$, the number $N$ of
clusters used, the age estimate and the relevant reference.}
\end{table}

Finally, we need an estimate of $t_{0}^{obs}$ to apply the method.
Following Rebolo et al. \cite{VSA}, one can choose $(t_{0}^{obs},
\sigma_t) = (14.4, 1.4) \ {\rm Gyr}$ as obtained by a combined
analysis of the WMAP and VSA data on the  CMBR anisotropy spectrum
and SDSS galaxy clustering. Actually, this estimate is model
dependent since Rebolo et al. \cite{VSA} implicitly assumes that
the $\Lambda$CDM model is the correct one. However, this value is
in perfect agreement with $t_{0}^{obs} = 12.6^{+3.4}_{-2.4} \ {\rm
Gyr}$ determined from globular clusters age \cite{Krauss} and
$t_{0}^{obs} > 12.5 \pm 3.5 \ {\rm Gyr}$ from radioisotopes
studies \cite{Cayrel}. For this reason, one is confident that no
systematic error is induced on the adopted  method using the
Rebolo et al. estimate for $t_{0}^{obs}$ even when testing
cosmological models other than the $\Lambda$CDM one.

\section{Testing the different classes of cosmological models}

The method  outlined above can be applied  to the dark energy
models described in Sect.\,2 in order to constrain their
parameters and see if they are viable candidates to explain cosmic
acceleration. To this aim, let us note that each one of the
presented models is roughly described by  few parameters which
are\footnote{We drop the subscript $0$ from $\Omega_{M,0}$ since
it does not give rise to any confusion here. Also, we use the
dimensionless parameter $h$ instead of the Hubble constant $H_0$.}
$(\Omega_M, h, w)$ for the QCDM model, $(q_0, h)$ for the
parametric density model and $(n, h)$ for the curvature
quintessence. For all the three classes of models, there is still
another parameter entering the fit, that is the delay factor $df$,
which we will marginalize over since it is not interesting for our
aims. Let us first consider the QCDM model. The main results are
plotted in Figs.\,\ref{fig: tauqcdm} and \ref{fig: clqcdm}. In the
first plot, we compare the estimated clusters age with the
quantity\,:

\begin{equation}
\tau(z) = t_L(z) + df
\label{eq: deftau}
\end{equation}
using the best fit values for the model parameters and  the delay
factor which turn out to be\,:

\begin{displaymath}
(\Omega_M, h, w) = (0.25, 0.70, -0.81) \ \ \ \ , \ \ \ \ df = 4.5 \ {\rm Gyr}
\end{displaymath}
giving $\chi^2 \simeq 0.04$. The $\chi^2$ value for the best fit
parameters (both for the QCDM model and the other dark energy
models considered) turns out to be quite small suggesting that
errors have been seriously overestimated. This is not surprising
given the  arbitrary way we have fixed the uncertainties on the
estimated age of the clusters. That this is likely to be the case
is also suggested by a qualitative argument. One could rescale the
errors on $t_{L}^{obs}(z_i)$ in such a way that $\chi^2 = 1$ for
the best fit model. Since for the best fit QCDM model, $\chi^2
\simeq \chi^2_{lt}$, this leads to multiply by almost $1/5$ the
uncertainties on $t_{L}^{obs}(z_i)$. If the error on $t_0$ were
negligible, this means that we should reduce the uncertainty on
the cluster age from 1 Gyr to 0.2 Gyr that is indeed a more
realistic value. The presence of an error on $t_0^{obs}$ slightly
complicates this argument, but does not change the main
conclusion. We are thus confident that the very low value of the
$\chi^2$ we get for the best fit model is only due to
overestimating the uncertainties on the clusters ages. However, we
do not perform any rescaling of the uncertainties since, to this
end, we should select a priori a model as the confidence one which
is contrary to the adopted philosophy by which we want to point
out the degeneracy. It is worth stressing, however, that such
rescaling does not affect anyway the main results.

\begin{figure}
\centering \resizebox{6.5cm}{!}{\includegraphics{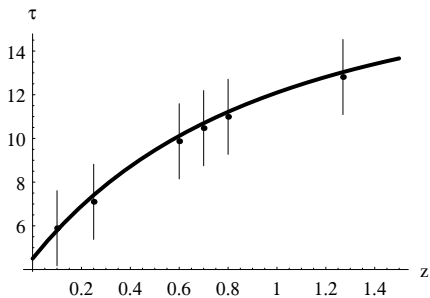}}
\hfill \caption{Comparison among predicted and observed values of
$\tau = t_L(z) + df$ for the best fit QCDM model.} \label{fig:
tauqcdm}
\end{figure}

\begin{figure}
\centering \resizebox{6.5cm}{!}{\includegraphics{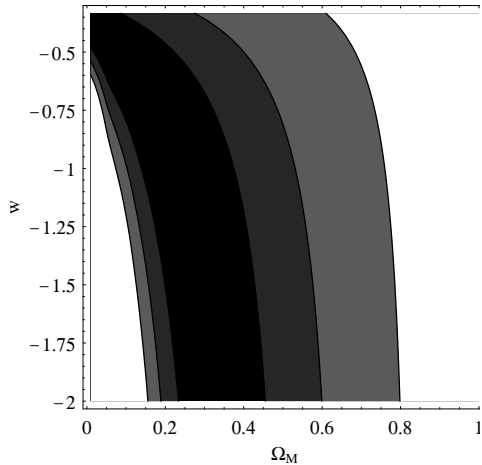}}
\hfill \caption{The $1 \sigma$ and $2 \sigma$ confidence regions
in the plane $(\Omega_M, w)$ for the QCDM model after
marginalizing over the Hubble constant $h$ and the delay factor
$df$.} \label{fig: clqcdm}
\end{figure}

\begin{figure}
\centering \resizebox{6.5cm}{!}{\includegraphics{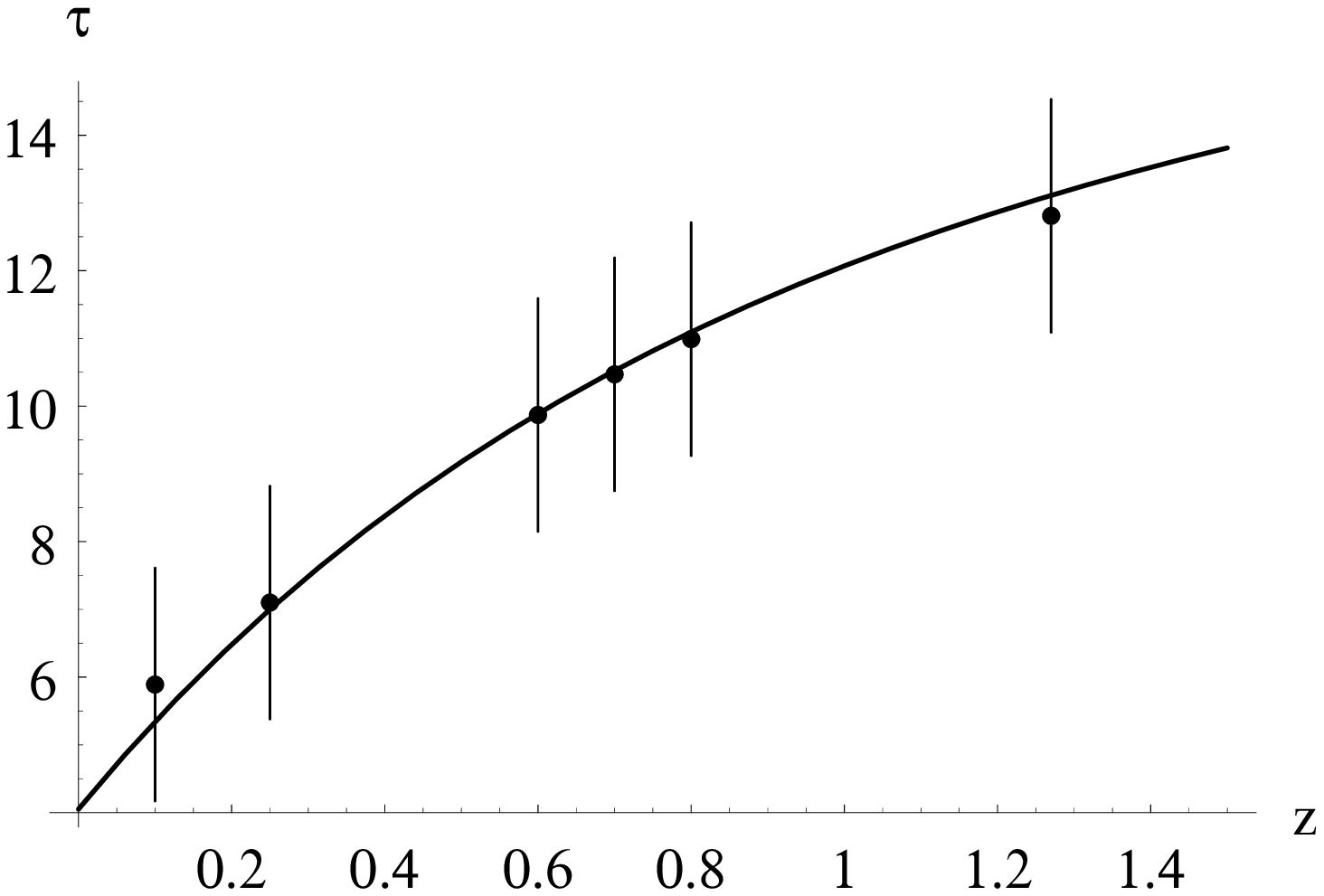}}
\hfill \caption{Comparison among predicted and observed values of
$\tau = t_L(z) + df$ for the best fit $\Lambda$CDM model.}
\label{fig: taulcdm}
\end{figure}

\begin{figure}
\centering \resizebox{6.5cm}{!}{\includegraphics{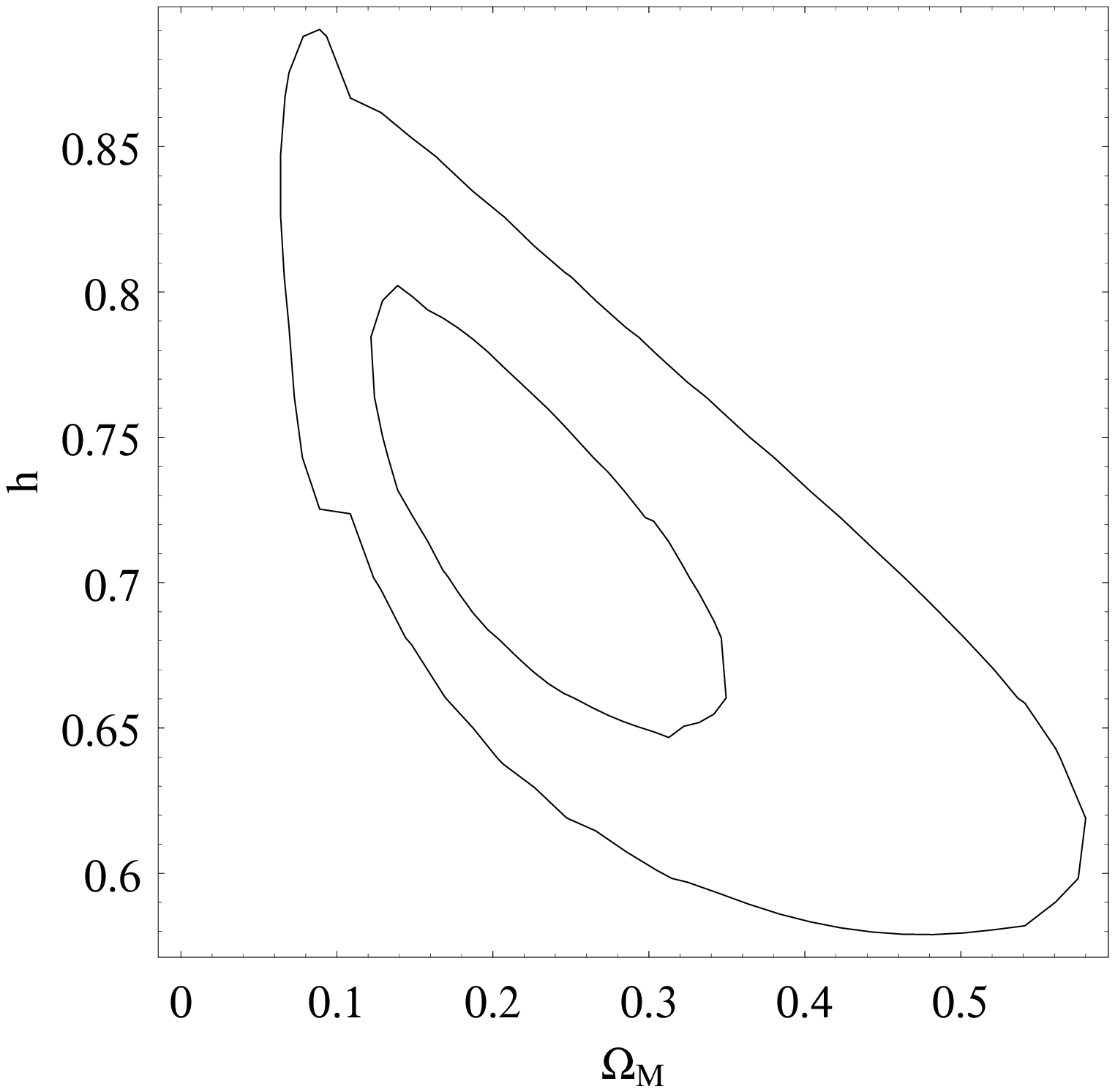}}
\hfill \caption{The $1 \sigma$ and $2 \sigma$ confidence regions
in the plane $(\Omega_M, h)$ for the $\Lambda$CDM model.}
\label{fig: cllcdm}
\end{figure}

Fig.\,\ref{fig: clqcdm} shows the $1 \sigma$ and $2 \sigma$
confidence levels in the $(\Omega_M, w)$ plane obtained by
marginalizing over the Hubble constant and the delay factor. Two
interesting considerations may be drawn from that plot. First, we
note that also {\it phantom} models (i.e. models with $w < -1$
violating the weak energy condition) are allowed by the data. This
is in agreement with recent results coming from fitting the QCDM
model to the SNeIa Hubble diagram and the CMBR anisotropy spectrum
\cite{Phantom}. Unfortunately, a direct comparison is not possible
since the marginalized likelihood is too flat to get any
constraints on $w$ so that all the values in the range tested $(-2
\le w \le 1/3)$ are well within the $1 \sigma$ CL. Although
essentially due to the paucity of the data, this result is also a
consequence of having not used any prior on $\Omega_M$ as it is
usually done in other analyses. By using the procedure described
at the end of Sect.\,3, we obtain the following estimates for the
other QCDM parameters\,:

\begin{displaymath}
\Omega_M \in (0.13, 0.39) \ \ \ \ , \ \ \ \ h \in (0.63, 0.77) \ \ \ \ (1 \sigma \ {\rm CL}) \ ,
\end{displaymath}

\begin{displaymath}
\Omega_M \in (0.01, 0.62) \ \ \ \ , \ \ \ \ h \in (0.56, 0.84) \ \ \ \ (2 \sigma \ {\rm CL}) \ .
\end{displaymath}
Given that we are able to give only weak constraints on the QCDM
model,  from now on I will  dedicate  attention to the case $w =
-1$, i.e. to the $\Lambda$CDM model and do not discuss anymore the
results for the QCDM model. The best fit parameters for the
cosmological constant model turn out to be\,:

\begin{displaymath}
(\Omega_M, h) = (0.22, 0.71) \ \ \ \ , \ \ \ \ df = 4.05 \ {\rm Gyr}  \ \ \ \ (\chi^2 \simeq 0.09)
\end{displaymath}
that gives rise to the curve $\tau(z)$ shown in Fig.\,\ref{fig:
taulcdm},  while Fig.\,\ref{fig: cllcdm} reports the confidence
regions in the $(\Omega_M, h)$ plane after marginalizing over the
delay factor. From the marginalized likelihood functions, we
get\,:

\begin{displaymath}
\Omega_M \in (0.10, 0.35) \ \ \ \ , \ \ \ \ h \in (0.63, 0.78) \ \ \ \ (1 \sigma \ {\rm CL}) \ ,
\end{displaymath}

\begin{displaymath}
\Omega_M \in (0.06, 0.59) \ \ \ \ , \ \ \ \ h \in (0.56, 0.85) \ \ \ \ (2 \sigma \ {\rm CL}) \ .
\end{displaymath}

The $\Lambda$CDM model has been widely tested against   a large
set of different astrophysical data. This offers us the
possibility to cross check both the model and the validity of
method. It is instructive, in this sense, to compare our results
with those coming from the fit to the SNeIa Hubble diagram. For
instance, Barris et al. \cite{Barris04} used a set of 120 SNeIa up
to $z = 1.03$ finding $\Omega_M = 0.33$ as best fit value with a
large uncertainty (not quoted explicitly, but easy to see in their
Fig.12) in good agreement with our result. A more recent result
has been obtained by Riess et al. \cite{Riess04} which have used a
SNeIa Hubble diagram extending up to $z = 1.55$ and have found
$\Omega_M = 0.29_{-0.03}^{+0.05}$ still in agreement with our
result. The $\Lambda$CDM model has also been tested by means of
the angular size\,-\,redshift relation. Using a catalog of
ultracompact radio sources and taking carefully into account
systematic uncertainties and selection effects, Jackson
\cite{Jack03} has found $\Omega_M = 0.24_{-0.07}^{+0.09}$ in
almost perfect agreement with our estimate.

\begin{figure}
\centering \resizebox{6.5cm}{!}{\includegraphics{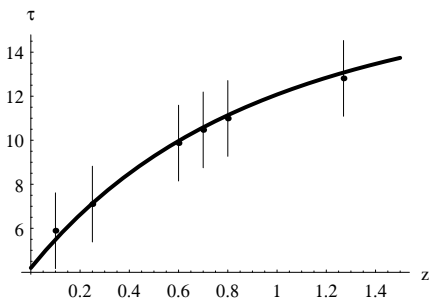}}
\hfill \caption{Comparison among predicted and observed  values of
$\tau = t_L(z) + df$ for the best fit parametric density model.}
\label{fig: tauhobbit}
\end{figure}

Neither Barris et al. \cite{Barris04} nor Riess et al.
\cite{Riess04} quote a best fit value for $h$ since this parameter
is infinitely degenerate with the supernovae absolute magnitude
$M$ when fitting to the SNeIa Hubble diagram. Nonetheless, SNeIa
may still be used to determine $h$ by fitting the linear Hubble
law to the low redshidt ($z < 0.1$) SNeIa. Using this method, Daly
\& Djorgovski \cite{DD04} have found $h = 0.664 \pm 0.08$ in good
agreement with the presented result. Moreover, it is worth noting
that our estimate for $h$ turns out to be in agreement with
estimates coming from different methods such as various local
standard candles \cite{Freedman}, the Sunyaev\,-\,Zel'dovich
effect in galaxy clusters \cite{SZ-Hz} and time delays in multiply
imaged quasars \cite{H0lens}. Finally, let us quote the results
obtained by Tegmark et al. \cite{SDSS03} which have performed a
combined fit of the $\Lambda$CDM model to both the WMAP data on
the CMBR anisotropy spectrum and the galaxy power spectrum
measured by more than 200,000 galaxies surveyed by the SDSS
collaboration. They find $\Omega_M = 0.30 \pm 0.04$ and $h =
0.70^{+0.04}_{-0.03}$ in very good agreement with our results.
Perhaps, the most interesting result of testing the $\Lambda$CDM
model with lookback  method is not the success of this model
(since it has yet been shown by a lot of evidences), but the
substantial agreement among these estimates of the parameters
$(\Omega_M, h)$ and those coming from distance measurements. This
is quite encouraging since it is an important successful cross
check  which, in general, could be applied to any cosmological
models.

\begin{figure}
\centering \resizebox{6.5cm}{!}{\includegraphics{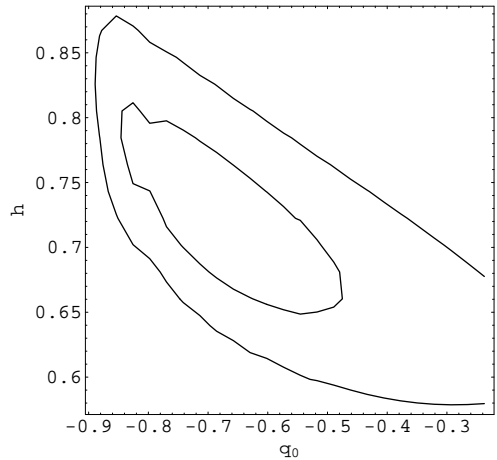}}
\hfill \caption{The $1 \sigma$ and $2 \sigma$ confidence regions
in the plane $(q_0, h)$ for the parametric density model.}
\label{fig: clhobbit}
\end{figure}

Let us now examine the results obtained for the parametric
density model, pictorially shown in Figs.\,\ref{fig: tauhobbit}
and \ref{fig: clhobbit}. The best fit model is obtained for the
following values of the parameters $(q_0, h)$ and of the delay
factor\,:

\begin{displaymath}
(q_0, h) = (-0.68, 0.71) \ \ \ \ , \ \ \ \ df = 4.20 \ {\rm Gyr}  \ \ \ \ (\chi^2 \simeq 0.07) \ .
\end{displaymath}
Marginalizing over $df$, we get\,:

\begin{displaymath}
q_0 \in (-0.81, -0.47) \ \ \ \ , \ \ \ \ h \in (0.64, 0.78) \ \ \ \ (1 \sigma \ {\rm CL}) \ ,
\end{displaymath}

\begin{displaymath}
q_0 \in (-0.89, -0.24) \ \ \ \ , \ \ \ \ h \in (0.58, 0.85) \ \ \ \ (2 \sigma \ {\rm CL}) \ .
\end{displaymath}
Note that the $2 \sigma \ {\rm CL}$ on the $q_0$  parameter  has
been truncated at the upper end since it formally extends to
values higher than the physically acceptable one.

In \cite{Hobbit}, the parameters of this model  have been
constrained by using both the SNeIa Hubble diagram and the angular
size\,-\,redshift relation. In particular, fitting the model to
the SNeIa Hubble diagram gives $h = 0.64_{-0.05}^{+0.08}$, while
the physically acceptable range for $q_0$ turns out to be in
agreement with the data for $q_0 = -0.42$ as best fit value. The
present day deceleration parameter $q_0$ is better constrained
using the data listed in Jackson \cite{Jack03} to perform the
angular size\,-\,redshift test thus obtaining $q_0 =
-0.64_{-0.12}^{+0.10}$ \cite{Hobbit}. Both these results are in
very good agreement with the present estimates so that we conclude
that the parametric density model is  a viable candidate
alternative to the $\Lambda$CDM model but this is a signal of
degeneracy among the two models.

\begin{figure}
\centering \resizebox{6.5cm}{!}{\includegraphics{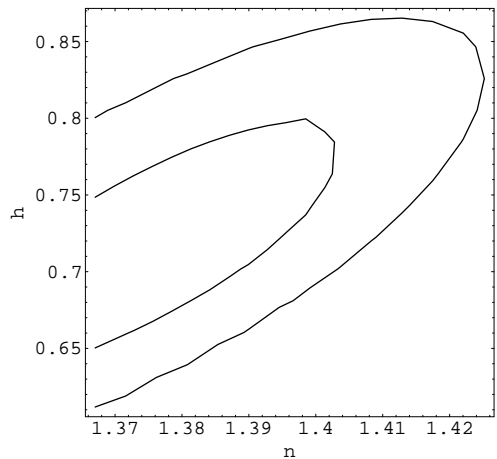}}
\hfill \caption{The $1 \sigma$ and $2 \sigma$ confidence regions
in the plane $(n, h)$ for curvature quintessence in the CurvUp
regime.} \label{fig: clcurvup}
\end{figure}

Finally, let us discuss the results  for the curvature
quintessence. Figs.\,\ref{fig: clcurvup} and \ref{fig: clcurvdown}
report the confidence regions in the plane $(n, h)$ for the CurvUp
and CurvDown regime respectively after marginalizing over the
delay factor. The first striking feature is that the contour plots
are not closed so that the marginalized likelihood function gives
only an upper (lower) limit to the parameter $n$ in the CurvUp
(CurvDown) regime. Formally, we get the following estimates for
the best fit values in the CurvUp and CurvDown regime
respectively\,:

\begin{displaymath}
(n, h) = (1.367, 0.71) \ \ \ \ , \ \ \ \ df = 4.80 \ {\rm Gyr}  \ \ \ \ (\chi^2 \simeq 0.23) \ ,
\end{displaymath}

\begin{displaymath}
(n, h) = (-0.367, 0.74) \ \ \ \ , \ \ \ \ df = 4.80 \ {\rm Gyr}  \ \ \ \ (\chi^2 \simeq 0.21) \ ,
\end{displaymath}
but the best fit value for $n$ actually  lies outside the
investigated range for this parameter. Being the confidence
regions open, it is meaningless to give constraints on $h$, but
nonetheless it is possible to infer the following limits\,:

\begin{figure}
\centering \resizebox{6.5cm}{!}{\includegraphics{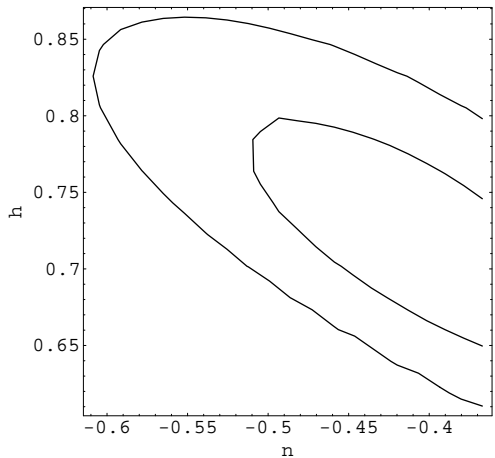}}
\hfill \caption{The $1 \sigma$ and $2 \sigma$ confidence regions
in the plane $(n, h)$ for curvature quintessence in the CurvDown
regime.} \label{fig: clcurvdown}
\end{figure}

\begin{displaymath}
n \le 1.402 \ \ {\rm at \ 1 \sigma \ CL} \ \ , \ \ n \le 1.424 \ \ {\rm at \ 2 \sigma \ CL} \ \ ,
\end{displaymath}

\begin{displaymath}
n \ge -0.508 \ \ {\rm at \ 1 \sigma \ CL} \ \ , \ \ n \ge -0.606 \ \ {\rm at \ 2 \sigma \ CL} \ \ ,
\end{displaymath}
where the first row refers to the CurvUp regime and  the second to
the CurvDown one. These limits do not contradict the ranges
determined by fitting to the SNeIa Hubble diagram \cite{curvfit},
but seems to be quite non-realistic. Actually, the deceleration
parameter corresponding to these values of $n$ is quite small
$(q_0 \sim -0.01)$ thus contradicting the evidence in favor of an
accelerating universe. Moreover, the results in \cite{curvfit}
have been obtained by using an old sample of SNeIa, including some
SNeIa that have now been discarded from the Gold set in Riess et
al. \cite{Riess04}. On the other hand, it is worth noting that
fitting a power law scale factor to the angular size\,-\,redshift
relation for compact radio sources gives $\alpha \simeq 1$
\cite{JDA03} which, by using Eq.(\ref{eq: alphancurv}), translates
in an estimate for $n$ that is in good agreement with the
presented result.

\section{Discussion and Conclusions}

The impressive amount of data   indicating a spatially flat
universe in accelerated expansion has posed the problem of dark
energy and stimulated the search for cosmological models able to
explain such unexpected behavior. Many rival theories have been
proposed to solve the puzzle of the nature of  dark energy ranging
from a rolling scalar field to a unified picture where a single
exotic fluid accounts for the whole dark sector (dark matter and
dark energy). Moreover, modifications of the gravity action has
also been advocated. Although deeply different in their underlying
physics, all these scenarios share the common feature of well
reproducing the available astrophysical data giving rise to a
degeneracy problem which is easily showed also in the
straightforward analysis presented here. It is worth stressing,
however, that the most widely used cosmological tests and in
particular the SNeIa Hubble diagram and the angular
size\,-\,redhisft relation) are essentially based on distance
measurements to high redshift objects and are thus affected by
similar systematic errors. It is hence particularly interesting to
look for  methods which are related to the estimates of different
quantities. Being affected by other kinds of observational
problems, such  methods could be considered as cross checks for
the results obtained by the usual tests and they should represent
complementary probes for cosmological models. The technique we
have devised here  is a first  step in this direction. We have
used the present age of the universe and the lookback time to
galaxy clusters to build a sort of {\it time diagram} of the
universe in order to reconstruct its age evolution. Relying on
stellar evolutionary codes, the estimate of the lookback time is
related to a different astrophysics than the distance based
methods and it is thus free from any problem connected with the
evolution of standard candles (such as the SNeIa absolute
magnitude and the intrinsic linear size of radio sources).
Actually, this technique could be affected by its own systematics
(such as, for instance, the degeneracy between age and
metallicity), but these may be better controlled. Moreover,
comparing the results thus obtained with those coming from
distance based methods allows to strengthen the conclusions
suggested by both techniques. Motivated by these considerations, I
have first applied the method to the $\Lambda$CDM model in order
to estimate the present day values of the matter density parameter
$\Omega_M$ and of the dimensionless Hubble constant $h$ obtaining
(at the $1 \sigma$ CL)\,:

\begin{displaymath}
\Omega_M = 0.22_{-0.10}^{+0.13} \ \ \ \ ,\ \ \ \ h = 0.71_{-0.08}^{+0.07} \ .
\end{displaymath}
These values are in agreement with the previous  estimates in
literature. This can be considered an independent confirmation not
only of the viability of the $\Lambda$CDM model, but also of the
method. It is worth noting that the $\Lambda$CDM scenario receives
further support from this test and henceforth the cosmological
constant $\Lambda$ still remains the best candidate to explain the
dark energy puzzle from an observational point of view.
Nonetheless, the $\Lambda$CDM model is severely  affected by
conceptual problems so that it is worthwhile to look at
alternative approaches. This has stimulated a plenty of models
where the cosmic acceleration is due to a dominant scalar field
rolling down its potential. However, such a scheme, on the one
hand, still does not solve the coincidence problem and, on the
other hand, is plagued by the unidentified nature of the scalar
field itself and the ignorance of its self interaction potential.
These considerations have opened the way to different models that
are able to give an accelerated expansion without the need of
scalar fields. Two of these approaches have been tested in these
lectures.

In \cite{Hobbit},  phenomenological unified models have been
proposed where a single fluid with a given energy density assigned
by few parameters is able to fit both the SNeIa Hubble diagram and
the angular size\,-\,redshift relation for ultracompact radio
sources.  These parametric density models have been tested here by
lookback time method and the following estimates (at the $1
\sigma$ CL) for its characterizing parameters have been
obtained\,:

\begin{displaymath}
q_0 = -0.68_{-0.13}^{+0.21} \ \ \ \ , \ \ \ \ h = 0.71 \pm 0.07 \ ,
\end{displaymath}
in agreement with previous tests \cite{Hobbit}. It is worth
stressing that such a method turns out to be more efficient than
the usual fit to the SNeIa Hubble diagram for the parametric
density models. It is also interesting to note that both $q_0$ and
$t_0$ predicted by these models are in almost perfect agreement
with those computed for the $\Lambda$CDM model with the parameters
$(\Omega_M, h)$ as discussed above. This is not surprising since
Eq.(\ref{eq: rhor}) shows that, nowadays, the energy density of
the parametric density model is very similar to that of the
$\Lambda$CDM model so that these two scenarios share most of the
observable properties referred to today quantities. However, this
does by no way mean that the two models are the same. Actually, it
is the underlying philosophy that is completely different having
now a single fluid rather than a cosmological constant dominating
over the matter term. Both models predict similar  values of the
today observed quantities  simply because they are tied to
reproduce the same data and not because they are two different
manifestations of the same underlying physics. Another possible
approach to the cosmic acceleration is  to consider dark energy
and dark matter as signals of the breakdown of the Einstein
General Relativity at some characteristic scale \cite{JCAP}. In
this picture, the universe is still dominated by  standard matter,
but the Friedmann equations have to be modified as consequence of
a different gravity action. This philosophy inspired curvature
quintessence scenarios where an effective dark energy is related
to the properties of the function $f(R)$ which replaces the Ricci
scalar $R$ in the gravity Lagrangian \cite{curvature}. The
coupling between matter and curvature for a theory with $f(R) =
f_0 R^n$ gives rise to fourth order nonlinear differential
equations for the evolution of the scale factor. Power law
solutions, $a \sim t^{\alpha}$, are possible, provided that
$\alpha$ is linked to $n$ by Eq.(\ref{eq: alphancurv}). Such
models are particularly attractive from a theoretical viewpoint
(since they allow to give a purely geometric interpretation of the
dark energy) and has also been shown to fit well SNeIa Hubble
diagram \cite{curvfit}. Here it is shown that, while they are able
to pass successfully the lookback time tests,
 only weak constraints can be imposed on the parameter $n$ with
the best fit value lying in a region corresponding to decelerating
rather than accelerating models. We are thus tempted to conclude
that this scenario could be rejected with some confidence.
Actually, this conclusion is also suggested by the  results in
Riess et al. \cite{Riess04} which, using the SNeIa Hubble diagram
up to $z = 1.55$, have  detected a change in the sign of the
deceleration parameter $q$. The power law solutions predict, on
the contrary, a constant $q$ so that they are ruled out by the
result in Riess et al. \cite{Riess04}. It is important to stress,
however, that this is not a definitive exclusion of the curvature
quintessence scenario. Actually, I have only considered simple
power law solutions. It is indeed conceivable that considering
more realistic models \cite{CNOT}, the evolution of the scale
factor gives a past deceleration followed by the present
acceleration thus leading the curvature quintessence models in
agreement with what is suggested by the extended SNeIa Hubble
diagram and other tests \cite{mimick}. Having tested three
different models,  it is worth asking what is the better one.
Unfortunately, this is not possible on the basis of the test
results only. Comparing the $\chi^2$ values, one could naively
conclude that the parametric density model is the best one since
it gives the lowest $\chi^2$. However, while the difference among
the $\chi^2$ values is significant between the parametric density
model and the curvature quintessence scenario (almost an order of
magnitude), it is too small (0.07 vs. 0.09) to conclude that the
$\Lambda$CDM model is disfavored. We have thus to conclude that
this test alone is not able to discriminate among these two dark
energy candidates. On the other hand, Sandvik et al.
\cite{Sandvik} recently claimed that UDE models are not viable
because the growth of density perturbations will lead to matter
power spectrum in disagreement with what is observed. This should
be an evidence against the parametric density model. However, it
is worth noting that Sandvik et al. explicitely consider the
generalized Chaplygin gas model which is characterized by a
negative squared sound speed that seems to be the main cause of
the anomalous growth of perturbations. For the parametric density
model, the sound speed is always positive definite so that it is
likely that the argument of Sandvik et al. should be at least
reconsidered. Finally, I would briefly comment on the possibility
to ameliorate  methods by increasing the maximum redshift probed.
To this aim, galaxy clusters do not appear as good candidates
since it is quite difficult to detect a significative number of
member galaxies up to redshifts larger than $\sim 1.3$. However,
high redshift galaxies may be taken into account provided that
they are detected in as more photometric bands as possible. This
latter requirement is fundamental since it allows not only to
better estimate the age of the galaxy, but also to infer
constraints on its formation redshift $z_F$ that cannot be assumed
to be the same for all the galaxies (as it has been possible for
clusters). Some techniques allow to discover galaxies up to
redshift $z \sim 6$ \cite{Dick04} and thus let us hope to measure
lookback time up to such high redshift. With its ability of both
furnishing multicolor photometry of high redshift galaxies (and
thus better estimates of their color ages and formation redshift)
and increasing the number of SNeIa, surveys like the GOODS one
\cite{GOODS} seems to be the most promising source of cosmological
constraints in the near future. In any case, from one hand we need
some {\it experimentum crucis} capable of removing the degeneracy
for a reasonably large redshift range and, from the theoretical
viewpoint, we need a physically reliable cosmological model,
emerging from some fundamental theory, without the conceptual
shortcomings of $\Lambda$CDM.

\section*{Acknowledgments}
I warmly thank G. Allemandi, A. Borowiec, V.F. Cardone, M.
Francaviglia, S. Nojiri, S.D. Odintsov, H.-J. Schmidt, T.
Sotiriou, A. Stabile and A. Troisi, for  comments, discussions and
suggestions on the presented topics.

\end{document}